\documentclass[twocolumn,aps,superscriptaddress,floats,prd,showpacs]{revtex4-2}

\usepackage{anyfontsize} 
\usepackage{amsmath,amssymb,amsfonts}
\usepackage{stmaryrd}
\usepackage{mathrsfs}
\usepackage{graphicx}
\usepackage{color} 

\begin{document}

\title{Solar-system experimental constraints on nonlocal gravity}

\author{Yunlong Liu}
\email[Corresponding author.]{phliuyunlong@kust.edu.cn}
\affiliation{School of Science, Kunming University of Science and Technology, Kunming 650093, China}

\author{Yongbin Du}
\affiliation{School of Physics, Sun Yat-sen University (Zhuhai Campus), Zhuhai 519000, China}

\date{\today}

\begin{abstract}
In this work, we study the constraints on the characteristic parameters
 $(\zeta,b)$  of the Deser-Woodard nonlocal gravity model in a static and
spherically symmetric background, using four classes of high–precision
Solar-System experiments: stellar light deflection, Shapiro time delay,
perihelion advance, and geodetic precession. From geodesic
equations, we derive observable geometric quantities that can be directly compared with VLBI/VLBA astrometry, the Cassini time–delay measurement, MESSENGER data and the GP-B/LLR results. 
Our results show that a larger value of $b$ suppresses the nonlocal
effect more rapidly with radius, thereby weakening the overall
constraints on $\zeta$. The perihelion advance exhibits the strongest
sensitivity to $\zeta$ around $b\simeq1.06$, providing the tightest
single experiment bound, whereas away from this region the combined
constraint becomes dominated by the Shapiro time delay. Incorporating
all four experiments yields a well–defined and sharply bounded
allowed region for the parameter space  $(\zeta,b)$.
\end{abstract}

\maketitle
\section{Introduction}
Over the past two decades, the precision of cosmological and gravitational experiments has increased dramatically. Observations ranging from supernova distance measurements \cite{Riess1998,Perlmutter1999}, anisotropies in the cosmic microwave background \cite{Planck2018,Planck2020} to the direct detection of gravitational waves \cite{GW150914,GW170817} have placed increasingly stringent demands on testing the validity of General Relativity (GR) across different physical scales.
In the weak field and Solar System regime, General Relativity has been stringently tested through high-precision PPN experiments \cite{Will2022}, including perihelion precession \cite{LAGEOS2010,Park2017}, light deflection \cite{Shapiro1964,Lebach1995,Shapiro2004,Fomalont2009}, gravitational redshift \cite{PoundRebka1960,Chou2010}, Shapiro time delay \cite{Bertotti2003}, and geodetic precession \cite{GPB2011,
Progress_Williams_2004}. In the strong field and cosmological regimes, GR is likewise supported by observations of binary pulsar energy loss \cite{HulseTaylor1975,TaylorWeisberg1982,Kramer2006}, black hole shadow \cite{EHT2019,Psaltis2020} measurements, gravitational wave merger waveforms \cite{LIGO_GRtest2019} and 
gravitational lensing \cite{Treu2010}. 
All these experiments demonstrate that General Relativity remains the most successful theory of gravity and spacetime to date.

Nevertheless, although General Relativity has passed high-precision tests across many regimes, extrapolating it beyond the experimentally verified scales, to either shorter ultraviolet (UV) distances or longer infrared (IR) distances, leads to tensions in both theoretical consistency and observational interpretation. 
At extremely small scales, the energy density is sufficiently high that the system inevitably collapses into a black hole, which has a singularity whose intrinsic curvature diverges, rendering existing physical theories inapplicable. To handle this problem, we need a quantum theory of gravity. However, attempts to construct a perturbative quantum theory of gravity that provides a self-consistent UV description are obstructed by the non-renormalizability of gravity \cite{tHooftVeltman1974,GoroffSagnotti1986}. Higher derivative extensions introduce to cure this non-renormalizable behavior, however, typically suffer from the appearance of ghost degrees of freedom \cite{Stelle1977,Woodard2015}. On the other hand, in order to account for the accelerated expansion of the Universe on large scales, General Relativity introduces a cosmological constant, which in turn leads to a conflict with the prediction of quantum zero-point energy \cite{Weinberg1989,Martin2012}. This fine-tuning problem uncovers that General Relativity also requires certain modifications in the infrared regime to provide a cosmological description that is consistent with quantum field theory.

Among the many attempts to modify gravity, nonlocal gravity stands out as particularly distinctive. While General Relativity is a manifestly local, differential theory, nonlocal gravity seeks to introduce modifications by incorporating terms involving the d'Alembert operator into the action, thereby implementing genuinely nonlocal extensions of GR. A well-known nonlocal strategy is the Deser-Woodard(DW) modification \cite{DeserWoodard2007,ParkWoodard2012,DeserWoodard2013,Woodard2014,Black_DAgostino_2025}, in which the Einstein-Hilbert action is supplemented by a term of the form $F(\Box^{-1}R)$. It therefore introduces a retarded kernel that can be fitted to yield an effective negative pressure at late times of the Universe, thereby accounting for the accelerated expansion. Because this retarded kernel arises from a global integral over the past causal domain, the theory is genuinely nonlocal. As a theory of gravity and spacetime, it does not require a dynamical explanation of $\Lambda$ to reproduce cosmic acceleration.
In this work, we use Solar-System gravitational experiments to constrain the parameters of this class of models. It is worth noting that, in addition to such infrared nonlocal modifications, there also exist nonlocal extensions aimed at addressing ultraviolet behavior \cite{Tomboulis1997,Biswas2012,Talaganis2015}.

High-precision tests of General Relativity often provide stringent bounds on the parameters of modified-gravity models. For example, Ref.~\cite{Brahma2021} considered a Kerr-like black hole solution incorporating loop quantum gravity (LQG) corrections. The LQG parameter modifies the radius of the photon ring, thereby affecting the angular diameter of the black hole shadow.
By comparing the theoretically predicted corrections with observational results from related measurements, highly precise constraints on the LQG parameter can be obtained. In Ref.~\cite{Liu2022}, Solar-System experiments were used to derive even tighter bounds, namely $0 < A < 4.0 \times 10^{-6}$.
In this work, following the same strategy, we neglect the extremely small rotational effects of Solar-System bodies and analyze Solar-System experiments within a static background to constrain the parameters of the DW nonlocal gravity model. The experiments considered include light deflection, Shapiro time delay, and the perihelion precession of Mercury. The observational data we employ are taken from the MESSENGER mission~\cite{Park2017} and LAGEOS II~\cite{LAGEOS2010}, whose analyses have already incorporated and removed rotational contributions.

This paper is organized as follows: In Sec.~\ref{BHSofNlG} and Sec.~\ref{EoMNlG}, we introduce the nonlocal gravity background and derived the corresponding equations of motion for test particles. 
In Sec.~\ref{OTNlG}, we employ numerical methods to obtain the constraints on the nonlocal parameters from four high–precision Solar System experiments respectively:
light deflection, Shapiro time delay, perihelion advance, and
geodetic precession. By combining these independent bounds, we extract the
tightest allowed region in the  $(\zeta,b)$  parameter space. Analytical
approximations supporting the numerical results are presented in the
Appendix.


\section{Black hole solution in nonlocal gravity} \label{BHSofNlG}
Following Ref.~\cite{DAgostino2025}, we define the auxiliary fields
\begin{align}
X  =  \Box^{-1}  R  & \Longleftrightarrow   \Box X  =  R ,  \\
Y  =  \Box^{-1} \left(g^{\mu\nu}\partial_{\mu}X\partial_{\nu}X\right)
&\Longleftrightarrow
\Box Y  =  g^{\mu\nu} \partial_{\mu}X   \partial_{\nu}X.
\end{align}
where $R$ is the Ricci scalar and $\Box = g^{\mu\nu}\nabla_{\mu}\nabla_{\nu}$. So the action of nonlocal gravity can be written as
\begin{align}
    S = \frac{1}{16\pi} \int \mathrm{d}^{4}x \sqrt{-g}  R \bigl( 1 + F[Y] \bigr) .
\end{align}
By introducing
\begin{align}
\Box U = -2 \nabla_{\mu}\left( V \nabla^{\mu} X \right) , 
\qquad
\Box V = R \frac{\mathrm{d}F}{\mathrm{d}Y}  ,
\end{align}
and use variation method , we can obtain the equations of motion
\begin{align}
\left( G_{\mu\nu} + g_{\mu\nu}\Box - \nabla_{\mu}\nabla_{\nu} \right) W
+\mathcal{K}_{(\mu\nu)}-
\frac{1}{2} g_{\mu\nu} g^{\alpha\beta}\mathcal{K}_{\alpha\beta}
= 0 ,
\label{EOM}
\end{align}
where $G_{\mu\nu} = R_{\mu\nu} - \frac{1}{2} g_{\mu\nu} R$ is Einstein tensor and $W = 1 + U + F[Y].$ Moreover,
\begin{align}
\mathcal{K}_{\mu\nu} = \partial_{\mu}X\partial_{\nu}U + \partial_{\mu}Y\partial_{\nu}V + V\partial_{\mu}X\partial_{\nu}X ,
\end{align}
and its symmetrization is $\mathcal{K}_{(\mu\nu)} = \frac{1}{2}\left( \mathcal{K}_{\mu\nu} + \mathcal{K}_{\nu\mu} \right).$
One may then solve Eq.\eqref{EOM} for a static and spherically symmetric metric of the form\cite{Black_DAgostino_2025}
\begin{align}\label{spt}
\mathrm{d}s^{2}
= - f[r]\mathrm{d}t^{2}+ \frac{1}{f[r]h[r]}\mathrm{d}r^{2}
+ g[r](\mathrm{d}\theta^2+\sin{\theta}\mathrm{d}\phi^2 ).
\end{align}
where
\begin{align}
f[r]&=1 - \frac{2M}{r}- \frac{\zeta M^{b}}{r^{b}} ,
\\[3mm]
h[r]&=\bigg(1 - \frac{2M}{r}+ \frac{\zeta M^{b+1}}
{3^{b} r^{b+1}\left( \frac{r}{M} - 3 \right)^{2}}
\nonumber\\
& \times \Big(3^{b}\frac{r}{M}(
b(\frac{r}{M}-3)(\frac{r}{M}-2)
+ 4\frac{r}{M} - 9)
\nonumber\\
&\qquad
- 3(\frac{r}{M}-2)(2\frac{r}{M}-3)
(\frac{r}{M})^{b}\Big) \bigg) \Big/ f[r] ,
\\[3mm]
g[r] &= r^{2}.
\end{align}
From the above expressions, one sees that $\zeta$ can be interpreted as a measure of the deviation of nonlocal gravity from General Relativity, and its magnitude should be much smaller than unity so that the prediction of GR is consistent with the experiments with high precision. In the limit $\zeta \to 0$, one has $F[Y]\to 0$, and the metric reduces to the Schwarzschild solution. The parameter $b$ controls the radial falloff of the modification and satisfies $b>1$. A larger value of $b$ implies that the nonlocal correction becomes increasingly short-ranged.

\section{Equations of motion of particles in nonlocal gravity black holes}\label{EoMNlG}
In this section, we extract the results of Ref.~\cite{Du2025} concerning the equations of motion in a static and spherically symmetric spacetime, since the metric considered therein has the similar functional form as the one adopted in this work.
For test particles moving in the spacetime of the form Eq.\eqref{spt}, the conserved quantities are given by
\begin{align}
C_{\xi}
= p^{a}\xi_{a}
+ \frac{1}{2} S^{ab}\nabla_{a}\xi_{b} ,
\end{align}
where $\xi$ is a Killing vector. Note that the Killing vectors are
$\xi^{a} = (\partial/\partial t)^{a}$ and
$\phi^{a} = (\partial/\partial \phi)^{a}$.
For non-spinning particles, i.e.\ for $s=0$, the conserved energy $E$ and angular momentum $J$
in the equatorial plane can be obtained as
\begin{align}
E = f[r] p^{t},\qquad
J = g[r] p^{\phi},
\end{align}
where $p^\mu=m dx^{\mu}/d\tau$ is four momentum. Without loss of generality, we introduce dimensionless parameters in place of the original ones:
\begin{align}
\tilde{E} = \frac{E}{m}, 
\qquad
\tilde{t} = \frac{t}{M},
\qquad
\tilde{r} = \frac{r}{M},
\qquad
\tilde{J} = \frac{J}{mM}.
\end{align}
For simplicity of expression, we omit the tilde on dimensionless parameters; for example, $J$ will represent $\tilde{J}$ in the following parts of the paper. Then we can obtain the equations of motion in the equatorial plane
\begin{align}\label{dphidtau}
\frac{\mathrm{d}t}{\mathrm{d}\tau} = \frac{E}{f[r]}, 
\qquad
\frac{\mathrm{d}\phi}{\mathrm{d}\tau} = \frac{J}{g[r]},
\end{align}

\begin{align}\label{drdtau}
\left( \frac{\mathrm{d}r}{\mathrm{d}\tau} \right)^{2}
= h[r]\left(
E^{2}
- f[r]\left( 1 + \frac{J^{2}}{g[r]} \right)
\right),
\end{align}
From the equations of motion, one can directly compute the light–deflection angle as well as the perihelion precession of bound orbits. These quantities will be analyzed in detail in the following sections.

\section{Observational tests in nonlocal gravity}\label{OTNlG}
	 
\subsection{Light deflection}
We begin by deriving the constraints on the model parameters from light–deflection measurements.
Combining Eqs.~\eqref{dphidtau} and \eqref{drdtau} and, for null trajectories, setting $\varepsilon=0$, the equations of motion on the $r-\phi$ plane reads
	\begin{align}
		\phi'[r]=\frac{d\phi}{d\tau}\frac{d\tau}{dr} = \frac{\sigma}{\sqrt{g[r]\left(-f[r]+b_{EJ}^2 g[r]\right)h[r]}}.
	\end{align}
where, $b_{EJ}=E/J$ and $\sigma=\mathrm{sgn}(dr/d\tau)=\pm 1$. 
On the ingoing branch one has $\sigma=-1$, whereas on the outgoing branch the sign reverses.
        The Deflection of light can be obtained as
	\begin{align}
		\Delta \phi =  2 \int_{r_0}^{\infty } \left|\phi'[r]\right|  dr-\pi,
	\end{align}
	Note that we have assumed that the light trajectory reaches its turning point at $r=r_{0}$, which corresponds to the point closest to the central mass. At this point, one has $\phi'[r]|_{r_{0}}=0$.
Then the parameter $b_{EJ}$ satisfies the following relationship
	\begin{align}
		b_{EJ}^2=f[r_0]/g[r_0]
	\end{align}
	
    By transforming $r = r_0/u$, we obtain the new expression for the deflection angle as:
	\begin{align}
		\phi'[u] &=\frac{\sigma r_0}{u^2\sqrt{g[r_0/u]\left(-f[r_0/u]+b_{EJ}^2 g[r_0/u]\right)h[r_0/u]}} \\
		\Delta \phi &=  2 \int_{0}^{1} \left|\phi'[u]\right| du-\pi,
		\label{DphiuI}
	\end{align} 
        For a more detailed analytical approximation solution, please refer to Appendix \ref{ASLD}. Here, we will use numerical methods for analysis.
        We assume that the deflection angle of starlight can be split into two parts:
	\begin{align}
		\Delta \phi = \Delta\phi_{GR}+\Delta\phi_{NlG} =\Delta\phi_{GR}\left( 1 + \delta_{NlG}\right),
	\end{align}
        Here, $\delta_{NlG}$ is the relative correction of the starlight deflection angle of nonlocal gravity with respect to gravity.
        In the numerical calculation, we use:
	\begin{align}
		\delta_{NlG}=\frac{\Delta \phi-\Delta \phi_{GR}}{\Delta \phi_{GR}},
	\end{align}
	
	\subsection{Shapiro time delay} 
    Next, we incorporate the Shapiro time–delay experiments and obtain the corresponding parameter constraints.
	Combining Eq.~\eqref{dphidtau} and \eqref{drdtau}, we can also acquire the equations of motion as
	\begin{align}\label{dt}
		t'[r] = \sigma \sqrt{  \frac{ b_{EJ}^2 g[r] }{ f[r]^2 \left( - f[r] + b_{EJ}^2 g[r] \right) h[r] } },
	\end{align}
    Therefore, the Schwarzschild coordinate time spent by the light ray when propagating from $r_{X}$ to $r_{0}$ is:
	\begin{align}
		\Delta t[r_X] = \int_{r_0}^{r_X} \left| t'[r] \right| dr,
		\label{DtrI}
	\end{align}
        We assume that an electromagnetic wave signal is emitted from the source point $X$ (satellite or Earth), and the signal reflected by the reflector $Y$ (satellite or planet) is received. When the electromagnetic wave signal passes through the perihelion, i.e., the superior conjunction cases, the following can be obtained.
	\begin{align}
		\delta t^S / 2 = \Delta t[r_X] +\Delta t[r_Y] -\left(\sqrt{r_X^2-r_0^2}+\sqrt{r_Y^2-r_0^2}\right)
	\end{align}
	 Here, the term $\delta t^S$ represents the Shapiro delay, while the last term $\sqrt{r_X^2-r_0^2}+\sqrt{r_Y^2-r_0^2}$ is the travel time that a light ray would spend following the same trajectory in a Minkowski spacetime.
	 Then the relative frequency change is
	 \begin{align}
	 	\delta \nu &= \frac{d}{dt} \delta t_S = \frac{d}{dt} \left( \delta t_S^{GR} +  \delta t_S^{NlG} \right)  \notag\\
	 	&\approx \delta \nu_S^{GR} + \frac{d}{dt} \delta t_S^{NlG} \approx \delta \nu_S^{GR} + \frac{d r_0}{d t} \frac{d \delta t_S^{NlG}}{dr_0}.
	 \end{align}
	Combining the experiments of Cassini \cite{Bertotti2003}, we can conclude that the frequency shift caused by modified gravity is
	\begin{align}
		\delta \nu_{NlG} \approx  \frac{d r_0}{d t} \frac{d \delta t_S^{NlG}[r_0]}{dr_0} < 10^{-14},
	\end{align}
	
	\subsection{Perihelion advance}
    In curved spacetime, a planetary orbit is no longer a closed ellipse but exhibits precession, which can be used to test gravity theory. Here, we compare our theoretical prediction with the MESSENGER mission \cite{Park2017} data to obtain the corresponding constraints on the nonlocal gravity parameter.
	Similarly, for timelike orbits, we can obtain the equations of motion as
	\begin{align}
		\phi'[r] = \sigma \sqrt{\frac{g[r]\left(E^2 g[r] + f[r]\left(J^2 + g[r]\right)\right)h[r]}{J^2}}
	\end{align}
	
        Assume that the perihelion and aphelion are $\{r_p, r_a\}$ respectively, and their relationship with energy $E$ and angular momentum $J$ is:
	\begin{align}
		E^2&= \frac{ f[r_p] f[r_a] \left(g[r_p]-g[r_a]\right)}{f[r_a] g[r_p] - f[r_p] g[r_a]}\\
		J^2&= \frac{ g[r_p] g[r_a] \left(f[r_p]-f[r_a]\right)}{f[r_a] g[r_p] - f[r_p] g[r_a]}
	\end{align}
	
\begin{figure}[!htb]
	\centering
        \includegraphics[width=0.45\textwidth]{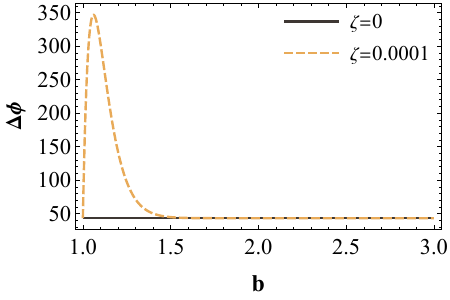}
		\caption{The angle of perihelion advance varying with $b$ for $\zeta=0.0001$
        Perihelion advance $\Delta\phi$ as a function of the parameter $b$ for the Schwarzschild limit ($\zeta=0$) and for a nonlocal gravity deformation ($\zeta = 0.0001$). The deformation significantly enhances the perihelion advance when $b$ is close to unity, producing a sharp peak around $b\simeq 1.06$, while for $b\gtrsim 1.2$ the effect becomes negligible and the two curves almost converges.
        }
		\label{DPhi}
\end{figure}

        The perihelion precession angle is
	\begin{align}
		\Delta \phi = 2 \int_{r_p}^{r_a}  \left|\phi'[r]\right| dr -2 \pi,
	\end{align}
	The numerical solution of the perihelion precession angle can be seen in Fig.~\ref{DPhi}. When $b \simeq 1.06$, the parameter $\zeta$ has the greatest impact on $\Delta \phi$. Thus, it will give strongest constraints on the parameter $\zeta$.
    
\subsection{Geodetic precession} 
Finally, we investigate the bounds on the modified–gravity parameter derived from geodetic precession measurements.        
    The equations of motion of a spinning particle can be described by the Mathisson-Papapetrou-Dixon(MPD) equation
    \begin{align}
\frac{D p^{a}}{D\tau} &= -\frac{1}{2}\, R^{a}{}_{bcd}\, v^{b} S^{cd} , \\[6pt]
\frac{D S^{ab}}{D\tau} &= - p^{b} v^{a} + p^{a} v^{b}.
\end{align}

    Considering the case where the spin is very small, in a weak field, the equations of motion of a spinning particle in a circular orbit can still be described by the circular orbit geodetic approximation equation as
    \begin{align}
        \frac{d t}{d\tau}&=\frac{E}{ f[r]}, \quad 
        \frac{d \phi}{d\tau}=\frac{J}{ g[r]} ,\\
        \frac{d r}{d\tau}&=\frac{d^2 r}{d\tau^2}=\frac{d \theta}{d\tau}=0,
    \end{align}
    where,  
    \begin{align}
        E &= \frac{ f[r] \sqrt{g'[r]}}{\sqrt{f[r] g'[r]-g[r] f'[r]}},\\
        J &= \frac{ g[r] \sqrt{f'[r]}}{\sqrt{f[r] g'[r]-g[r] f'[r]}}.
    \end{align}

    We have:
        \begin{align}
        \frac{ds^t}{d t} &= -\frac{f'[r] s^r}{2 f[r]},\\
        \frac{ds^r}{d t} &= \frac{ f[r] h[r] \left( g'[r] \dot{\phi } s^\phi-f'[r] \dot{t } s^t\right)}{2 \dot{t}},\\
        \frac{ds^\theta}{d t} &=0 ,\\
        \frac{ds^\phi}{d t} &= -\frac{ g'[r] \dot{\phi } s^r}{2  g[r] \dot{t}},
        \end{align}
    Furthermore, we can obtain
    \begin{align}
        \frac{d^2 s^r}{d t^2} = -\omega^2 s^r
    \end{align}
    where, 
    \begin{align}
        \Omega^2&= \frac{\dot{\phi}}{\dot{t}} =\frac{f'[r]}{g'[r]}\\
        \omega^2&=   \frac{1}{4} h[r] f'[r] g'[r] \left(\frac{f[r]}{g[r]}-\Omega^2\right)
\end{align}

Consider initial conditions: 
\begin{align}
	\label{Ins}
	s^r|_{t=0}=s^r_0\neq 0,  \quad  {s^r}'|_{t=0}=0,
	\\ s^t|_{t=0} = s^{\theta}|_{t=0} = s^{\phi}|_{t=0} = 0.
\end{align}

We can obtain the solution 
\begin{align}
	s^t=&-\frac{f'[r]}{2 f[r] \omega}\sin [ \omega t] s^r_0 ,\\
	s^r =& \cos [\omega t] s^r_0 ,\\
	s^{\theta }=&0,\\
	s^{\phi }=&-\frac{\Omega g'[r] }{2 \omega g[r] }\sin[ \omega t] s^r_0.
\end{align}
In the formula, $\omega$ can be interpreted as the rotational velocity of the spin vector $s^a$. The ratio of the rotational velocity $\omega$ to the angular velocity $\Omega$ of the orbit is as follows:
\begin{align}
	\frac{\omega^2}{\Omega^2}
	=\frac{h[r] g'[r] \left(f[r] g'[r]-g[r] f'[r]\right)}{4 g[r]}.
\end{align}
For one period of motion, the geodetic precession angle can be expressed as
\begin{align}
	\label{Geop}
	\Delta \Theta = 2 \pi (1-\frac{\omega}{\Omega})
\end{align}

	\subsection{Total constraint}
\begin{figure}[!htb]
		\centering
		\includegraphics[width=0.45\textwidth]{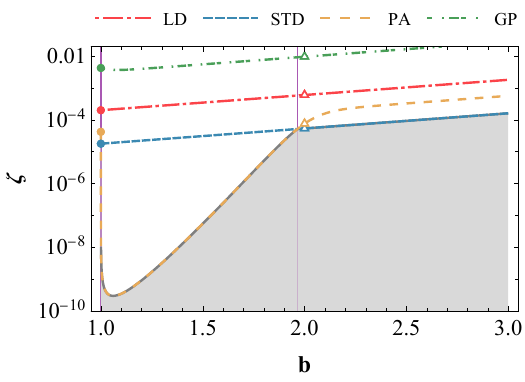}
		\caption{The total constraint of the parameters $(\zeta,b)$  obtained from the four tests considered in this work. The bounds derived from the light deflection(LD), Shapiro time delay(STD), perihelion advance(PA) and geodetic precession(GP) are show by the red, green, orange and bule dashed line, respectively. The shaded gray region indicates the parameter space bounded by all experiments.}
		\label{CTALL}
	\end{figure}

    Using data from  VLBA\cite{Fomalont2009}, Cassini\cite{Bertotti2003}, MESSENGER mission\cite{Park2017} and Gravity Probe B\cite{GPB2011}, 
    the constraints on $\zeta$ can be obtained and presented together in a single figure \ref{CTALL}  for different values of $b$. This allows a direct comparison of the sensitivity of light deflection, Shapiro time delay, perihelion precession, and geodetic precession measurements to the nonlocal gravity parameter. Note that the solid dots and hollow triangles in the figure represent the constraint ranges obtained through the analytical approximation method when $b=1$ and $b=2$, respectively. Detailed comparisons are provided in Appendix \ref{AS}. It can be seen that the results from the numerical method and the analytical method are consistent.
    
    We can see that as $b$ increases, the ability to constrain $\zeta$ becomes increasingly weaker, in general. 
    However, when $1 < b < 2$, some anomalous phenomena occur.
    For example, for the PA constraint, the tightest bound $|\zeta| \lesssim 3\times 10^{-10}$ is at $b \simeq 1.06$. 
    
    Specifically, the two vertical lines represent $b=1.0000002246$
    and $b=1.9645$, respectively. This means that within the range of $[1.0000002246, 1.9645]$, the perihelion precession experiment has a stronger constraining ability on $\zeta$. In the ranges of $[1, 1.0000002246]$ and $[1.9645, 3]$, the time delay experiment has a stronger constraining ability on $\zeta$.
    Overall, the combined analysis significantly narrows the allowed region of  $(\zeta,b)$, demonstrating that Solar System experiments place stringent limits on the nonlocal gravity corrections.

\section{Conclusions}
Starting from DW nonlocal gravity model on a static and
spherically symmetric background, the deviation from GR of which is characterized as the parameter $b$ and $\zeta$, we consider four classes of Solar System observations, including light deflection, Shapiro time delay, perihelion precession, and geodetic precession, to give a tight constraints on the value of the parameter. Together with the analytical approximations evaluated at $b=1$ and $b=2$, our numerical analysis provides a coherent and systematic constraint on the model parameters.
Our results show that, as $\zeta\to 0$, all observational tests naturally recover the Schwarzschild limit, and the combined bounds significantly shrink the allowed region of $\zeta$. 
Meanwhile, the scale parameter $b$ controls the radial decay
of the nonlocal modification; larger $b$ often corresponds to a larger bound on $\zeta$.
Notably, the perihelion advance exhibits a sharp response near $b\simeq1.06$, providing the strongest individual bound across all experiments, whereas the Shapiro delay sets the most sensitive constraint when $b$ is large. 
The complementarity among these tests ensures that the final combined constraint is both stable and robust.

Looking ahead, the most natural extension of this work is to generalize
the static background to include slow rotation, enabling joint
constraints from orbital dynamics and Lense-Thirring precession.
Furthermore, future high-precision Solar System missions will continue
to improve the sensitivity to nonlocal effects. On the one hand,
observations already show no detectable deviation from general
relativity, offering a clear, operational pathway for testing nonlocal
gravity in the high-precision regime. On the other hand, the parameter
space  $(\zeta,b)$  has now been sharply delineated, providing concrete
targets for future experimental refinement.

\section*{Acknowledgements}
We thank our colleagues for valuable discussions on nonlocal operators and radio-science systematics.

\appendix
	
	\section{Analytical solution}\label{AS}
	
	\subsection{Light deflection}\label{ASLD} 
	\subsubsection{For b=1}
	When $b=1$, the equations of motion can be expanded approximately as 
	\begin{align}
		\phi'[u]&=\frac{1}{\sqrt{1 - u^2}} + \frac{(2 + \zeta)  (1 + u + u^2)}{2 r_0 (1 + u) \sqrt{1 - u^2}} \notag\\
		&+ \frac{3 (2 + \zeta)^2 (1 + u + u^2)^2}{8 r_0^2 (1 + u)^2 \sqrt{1 - u^2}}+\mathcal{O}[\left(\frac{1}{r_0}\right)^3]
	\end{align}
        Considering Eq.\eqref{DphiuI}, by integrating each term, we can obtain the results of the first two orders.
        Then, we can get
	\begin{align}
		\Delta \phi \approx \frac{4 }{r_0}\left(1+\frac{\zeta}{2}\right)\approx \Delta \phi_{GR}\left(1+\frac{\zeta}{2}\right)
	\end{align} 
    
        Combined with the experimental data\cite{Fomalont2009}, the constraint range of $\zeta$ can be obtained as
	\begin{align}
		-0.001<\zeta<0.0002
	\end{align}
    
	\subsubsection{b=2}
        When $b = 2$, the equations of motion can be expanded as
	\begin{align}
		\phi'[u]&=
		\frac{1}{\sqrt{1 - u^2}} + \frac{ \left( 3 + (3 + \zeta) u (1 + u) \right)}{3 r_0 \sqrt{1 - u}  (1 + u)^{3/2}} 
		+\mathcal{O}\left[\frac{1}{r_0^2}\right]
	\end{align}
        Consider Eq.\eqref{DphiuI}, integrating each term, we can obtain the results of the first two stages. From this, we can get
	\begin{align}
		\Delta \phi \approx \frac{4 }{r_0}\left(1+\frac{\zeta}{6}\right)\approx \Delta \phi_{GR}\left(1+\frac{\zeta}{6}\right)
	\end{align}
	With the data in Ref.~\cite{Fomalont2009}, the constraint range of $\zeta$ can be obtained.
	\begin{align}
		-0.003<\zeta<0.0006
	\end{align}
	
	\subsection{Shapiro time delay}\label{ASSTD} 
	\subsubsection{For b=1}
        When $b=1$, the Eq.\eqref{dt} can be approximately expanded as
	\begin{align}
		t'[u]&=
		\frac{r_0}{u^2 \sqrt{1 - u^2}} + \frac{(2 + \zeta)  (2 + 3 u)}{2 u (1 + u) \sqrt{1 - u^2}} \notag\\
		&+ \frac{3 (2 + \zeta)^2  (4 + 8 u + 5 u^2)}{8 r_0 (1 + u)^2 \sqrt{1 - u^2}}
		+\mathcal{O}[\frac{1}{r_0^2}]
	\end{align}
        Considering Eq.\eqref{DtrI}, integrating each term, we can obtain the results of the first two orders. Finally, the approximate solution of Shapiro time delay can be derived as
	\begin{align}
		\Delta t &\approx  \left(1+\frac{\zeta}{2}\right)   \left(4 +  4 \ln[\frac{4 r_X r_Y}{r_0^2}] - \frac{2 r_0\left(r_X+r_Y\right)}{r_X r_Y}\right) \notag\\
		&\approx \Delta t_{GR}\left(1+\frac{\zeta}{2}\right)
	\end{align}
        Take the derivative of $r_0[t]$
	\begin{align}
		\Delta \nu &\approx 
		- \left(1+\frac{\zeta}{2}\right) \frac{1}{r_0} \left(8+\frac{2 r_0 (r_X+r_Y)}{r_X r_Y} \right)  r_0'[t]
		 \notag\\
		&\approx \Delta \nu_{GR}\left(1+\frac{\zeta}{2}\right)
	\end{align}
    
	Combined with the experimental data, the constraint range of $\zeta$ can be obtained
	\begin{align}
		-1.8 \times 10^{-5}<\zeta<1.8 \times 10^{-5}
	\end{align}
	
	\subsubsection{b=2}
        When $b = 2$, the equations of motion can also be expanded as
	\begin{align}
		t'[u]&=
		\frac{r_0}{u^2 \sqrt{1 - u^2}} + \frac{  \left(6 + 9 u + \zeta(1+u) \right)}{3 u (1 + u) \sqrt{1 - u^2}} +\mathcal{O}[\frac{1}{r_0^2}]
	\end{align}
    Consider Eq.\eqref{DtrI}, integrate each term, and we can obtain the results of the first two stages. Finally, the approximate solution can be obtained as
	\begin{align}
		\Delta t &\approx  M  \left(4 +  \left(4+\frac{2\zeta}{3}\right) \ln[\frac{4 r_X r_Y}{r_0^2}] - \frac{2 r_0\left(r_X+r_Y\right)}{r_X r_Y}\right) \notag\\
		&\approx \Delta t_{GR}  +  \frac{2\zeta}{3}\ln[\frac{4 r_X r_Y}{r_0^2}] 
	\end{align}

        Similarly, take the derivative of $r_0[t]$
	\begin{align}
		\Delta \nu &\approx 
		- \frac{1}{r_0} \left( 4 + \frac{2 \zeta}{3}+\frac{r_0 (r_X+r_Y)}{r_X r_Y} \right)  r_0'[t]
		\notag\\
		&\approx \Delta \nu_{GR} - \frac{2 \zeta }{3 r_0}
	\end{align}
	
	Combined with the experimental data, the constraint range of $\zeta$ can be obtained
	\begin{align}
		-5.3 \times 10^{-5}<\zeta<5.3 \times 10^{-5}
	\end{align}

	\subsection{Perihelion advance}\label{ASPA} 
	\subsubsection{b=1}
	When $b=1$, the equations of motion can be expanded approximately as
	\begin{align}
		&u''[\phi] + u[\phi] =K_0+ \frac{3 (2+\zeta)   u [\phi]^2}{2 r_0} \label{dduphi} \\
		&K_0 = \frac{(\zeta+2)  r_0}{2 J^2}  
	\end{align}
     We use the perturbation method to calculate the analytical solution of the precession angle approximately. For the leading order, the equation $u_0''[\phi] + u_0[\phi] = K_0$ is satisfied. It is assumed that the solution of $u_0[\phi]$ at the perihelion satisfies the approximate solutions $u_0[0] = K_0(1 + e)$ and $u_0'[0] = 0$, where $e$ is the eccentricity. The leading-order approximate solution can be obtained as follows:
	\begin{align}\label{u0Soution}
		u_0[\phi] =K_0(1+e \cos[\phi]) 
	\end{align}
        We consider the first-order approximation, and consider the correction as $u[\phi] \approx u_0[\phi] + u_1[\phi]$. Substituting into equation \eqref{dduphi}, we can obtain that the approximate solution satisfies
	\begin{align}\label{u1Soution}
		u_1''[\phi] + u_1[\phi] = \mathcal{A}_0 +\mathcal{A}_1 \cos[\phi]+\mathcal{A}_2 \cos[\phi]^2
	\end{align}
	where,
	\begin{align}
		\mathcal{A}_0 &= \frac{3(2+\zeta) K_0^2 }{2 r_0} \\
		\mathcal{A}_1 &= \frac{3(2+\zeta) e K_0^2 }{r_0} \\
		\mathcal{A}_2 &= \frac{3(2+\zeta) e^2 K_0^2 }{2 r_0}
	\end{align}

        The approximate solution can be obtained as
	\begin{align}
		&u_1[\phi] =  \frac{1}{2} \mathcal{A}_1 \phi \sin[\phi] \notag\\
		&+  \mathcal{A}_0 + \frac{\mathcal{A}_2}{2} - \frac{1}{3} (3\mathcal{A}_0 + \mathcal{A}_2) \cos[\phi] - \frac{1}{6} \mathcal{A}_2 \cos[2\phi]
	\end{align}
        Only $\sin[\phi]$ has an effect on the precession angle, so by only considering the $\sin[\phi]$ term, the approximate solution for $u[\phi]$ can be obtained as
	\begin{align}
		u[\phi] \approx K_0 + e K_0 \cos [\phi]+ \frac{1}{2} \mathcal{A}_1 \phi  \sin[\phi]
	\end{align}
	
        Finally, we can obtain the approximate deflection angle of perihelion precession.
	\begin{align}
		\delta \phi_0 \approx \left(1+\frac{\zeta}{2}\right)\frac{6 K_0  \pi}{r_0} \approx \left(1+\frac{\zeta}{2}\right)\delta \phi_0^{GR}
	\end{align}
        Combined with the precession error of Mercury's perihelion, we can obtain $-4.188 \times 10^{-5} < \zeta < 4.188 \times 10^{-5}$.
	
	\subsubsection{b=2}
	When $b=2$, the equations of motion can be expanded approximately as
	\begin{align}
		u''[\phi] + u[\phi] =& K_0-\frac{\zeta   u[\phi]}{3 J^2} \notag\\
		&+\frac{ u[\phi]^2 \left((\zeta +3) J^2-E^2 \zeta ^2 \right)}{J^2 r_0}  \label{dduphib2} 
	\end{align}
	where, 
	\begin{align}
		K_0 =  \frac{ r_0 \left(3 + \zeta  \left(1-E^2\right)\right)}{3 J^2}.
	\end{align}
    Similar to Eqs.~\eqref{u0Soution} - \eqref{u1Soution}, we can re-obtain
	\begin{align}
		\mathcal{A}_0 &= \frac{K_0  \left(3 K_0 \left((\zeta +3) J^2-E^2 \zeta ^2 \right)-\zeta   r_0\right)}{3 J^2 r_0} \\
		\mathcal{A}_1 &= \frac{e K_0  \left(6 (\zeta +3) K_0 J^2-6 E^2 \zeta ^2 K_0 -\zeta   r_0\right)}{3 J^2 r_0}\\
		\mathcal{A}_2 &=\frac{e^2 K_0^2  \left((\zeta +3) J^2-E^2 \zeta ^2 \right)}{J^2 r_0}
	\end{align}

    Similarly, by referring to the calculation process for $b=1$, we can obtain the approximate deflection angle of perihelion precession  as
	\begin{align}
		\delta \phi_0 \approx \frac{\pi   \left(6 (\zeta +3) K_0 J^2-6 E^2 \zeta ^2 K_0 -\zeta  r_0\right)}{3 J^2 r_0}
	\end{align}
        Combined with the Mercury perihelion precession experiment, we can obtain $-7.539 \times 10^{-5} < \zeta < 7.539 \times 10^{-5}$.


\end{document}